\documentclass[fleqn,usenatbib]{mnras}

\usepackage{newtxtext,newtxmath}

\usepackage[T1]{fontenc}
\usepackage{ae,aecompl}

\usepackage{graphicx}	
\usepackage{amsmath}	
\usepackage{amssymb}	

\title[]{Hydrogen  spin oscillations in a background of axions and the 21-cm brightness temperature}

\author[Lambiase and Mohanty]{
G. Lambiase,$^{1,2}$\thanks{E-mail: lambiase@sa.infn.it}
S. Mohanty,$^{3}$
\\
$^{1}$Dipartimento di Fisica E.R. Caianiello Universit\`a di Salerno, I-84084 Fisciano (Sa), Italy\\
$^{2}$INFN - Gruppo Collegato di Salerno, Italy\\
$^{3}$Physical Research Laboratory, Ahmedabad 380009, India
}

\def\be{\begin{equation}}
\def\ee{\end{equation}}

\def\bea{\begin{eqnarray}}
\def\eea{\end{eqnarray}}

\date{Accepted XXX. Received YYY; in original form ZZZ}

\pubyear{2015}

\begin{document}
\label{firstpage}
\pagerange{\pageref{firstpage}--\pageref{lastpage}}
\maketitle

\begin{abstract}
The 21-cm line signal arising from the hyperfine interaction in hydrogen has an important role in cosmology and provides a unique method for probing of the universe  prior to the star formation era.
We propose that  the spin flip of Hydrogen by the  coherent emission/absorption of axions  causes a lowering of their spin temperature and can explain the stronger than expected  absorption of 21-cm light reported by the  EDGES collaboration. We find the analogy of axion interaction with the two level HI with the Jaynes-Cummings model of a two level atom in a cavity and we derive the spin flip frequency in this formalism and show that the coherent oscillations frequency $\Omega \propto 1/f_a$ in contrast with the incoherent transitions between the HI hyperfine levels where the transition rates $\propto 1/f_a^2$. The axion  emission and absorption  rates are equal but the spin temperature is still lowered due to different selection rules for the spin flip transitions compared to the photon process. We show that the axion process goes in the right direction for explaining the EDGES observation.
For this mechanism to work we require a coherent field of relativistic axions with energy $E_\nu$ peaked at the 21-cm spin-flip energy. Such a coherent background of relativistic axions can arise from the decay of cosmic strings if the decay takes place in the electroweak era.

\end{abstract}



\section{Introduction}

The first detection of the 21-cm signal from the era of first star formation called Cosmic Dawn ($15<z<35$)  \citep{Furlanetto:2006jb,Madau:1996cs} was recently reported by the EDGES experiment \citep{Bowman:2018yin}. Photons emitted by the spin flip from ortho (J=1) to para (J=0) hydrogen has energy $E_{21}= 5.9 \mu {\rm eV}$ corresponding to the frequency 1420  MHz from hyperfine interaction . When the CMB passes through a HI cloud at redshift $z$ there will be a dip observed in the CMB spectrum at frequency $1420 /(z+1)$ MHz  due to $J=0\rightarrow 1$ transitions. The  EDGES experiment observed a dip in the CMB frequency range 50-100 MHz centered at 78 MHz (corresponding to the redshift z=17.3)  and the brightness temperature $T_{21}= -500^{+200}_{-500} $ mK whereas the lower limit  from standard cosmology  is -209 mK (at $z\sim 17 $). The explanation for this anomalous dip which has been offered is that the gas cools by interaction with dark matter which in turn lowers the spin temperature (which is defined by the ratio of the para and ortho hydrogen states) \citep{Barkana:2018lgd,Fialkov:2018xre}. The velocity of the gas is lowest in the cosmic dawn era and a dark matter-baryon cross section $\sigma(v) =\sigma_1 (v/1 km s^{-1})^{-4}$ with $\sigma_1= 8 \times 10^{-20} cm^2$ and  dark matter mass of 0.3 GeV gives the best fit to the EDGES observation \citep{Barkana:2018lgd}. The $v^{-4}$ velocity dependence can come from the exchange of massless mediators. There are strong constraints from fifth force searches on light particle mediators  and the simplest possibility is baryon interaction with a milli-charged dark matter by photon exchange \citep{Munoz:2018pzp,Barkana:2018qrx}. However, if constraints from CMB distortion by dark matter annihilation \citep{Dvorkin:2013cea,Fialkov:2018xre,Berlin:2018sjs,DAmico:2018sxd} are taken into account the parameter space of milli-charged dark matter is $m_\chi = 10-80$ MeV, electric charge $\epsilon =10^{-6}-10^{-4}$ and a relic abundance which is a small fraction (0.3-2)\% of the dark matter \citep{Munoz:2018pzp, Berlin:2018sjs}.
Acoustic oscillations of the milli-charged dark matter would be observable in anisotropies of the 21-cm temperature power spectrum which would be observable with LOFAR and HERA \citep{Munoz:2018jwq}.
Other explanations for explaining the excess dip in the 21-cm absorption is conversion of 21-cm photons to dark photons \citep{Pospelov:2018kdh}. The possibility that the anomalous dip is due to an excess of background CMB photons in the $E_{21}$ wavelengths arising from the decay of photo-philic light dark matter has been studied \citep{Fraser:2018acy}. The cooling of hydrogen gas by 0.15 eV  QCD axion  dark matter has been proposed \citep{Houston:2018vrf, Houston:2018vbk} and the creation of excess photon at 21-cm by axion-photon conversion in cosmological magnetic field has been studied \citep{Moroi:2018vci}.

The spin temperature of Hydrogen can also change by a spin-flip absorption and emission of axions by  by either the electron or the proton in HI. In ref. \citep{Auriol:2018ovo} the rate for hydrogen spin-flip by axion absorption and emission was calculated and it was concluded that (1) spin-flip by axion emission and absorption from the background {\it always} increases the spin temperature of HI, so this mechanism goes in the wrong direction if one is attempting to explain the EDGES anomaly and (2) the rate of axion induced processes  smaller than the photon processes by  18 or so orders of magnitude.

In this paper we explain this discrepancy as due to the lowering of the spin temperature by a spin-flip of the electron or proton in HI by coherent absorption/emission of axions from the  background  of an axion field in a coherent state. This  spin oscillation frequency goes as $\Omega \propto 1/f_a$  ( $f_a$ being the axion decay constant)  We show that the HI spin oscillations in an axion bath is analogous to the Jaynes-Cummings model of a two level atom in a cavity. Therefore in contrast to the incoherent process where the axion-induced spin transition rate $ \Gamma \propto 1/f_a^2$, the axion induced coherent oscillations of the HI between the singlet and triplets states is much larger than the incoherent absorption/emission studied in \citep{Auriol:2018ovo}.  We also show that the spin temperature of HI is {\it always} lowered by the coherent emission and absorbtion of axions. This is so because the axilon induced spin flip occurs between the singlet state $|J-=0,M=0\rangle $ and one linear combination of the three triplet states $|J=0, M=0,\pm 1 \rangle$. This is unlike the photon process where spin transitions take place between the singlet and all three of the triplet states. For this mechanism to work without unnaturally fine tuning the axion mass to the 21-cm energy to many decimal places , the coherent axion  background must be have a large occupancy in the state with momentum $p_a= \sqrt{E_{21}^2 - m_a^2}$. Such an axion can arise from the decay of cosmic strings. The peak frequency of the axions in such a scenario is related to the time of the string decay as $k = \epsilon /(2 \pi t)$ with $\epsilon =0.25 \pm 0.18$ \citep{Yamaguchi:1998gx,Hiramatsu:2010yu,Gorghetto:2018myk}.


\section{Axion induced spin flip  contribution to spin temperature}

The neutral hydrogen makes  $ J=1\rightarrow 0 $ transitions due to axion emission which lowers the spin temperature at equilibrium with gas and photons.  In the  detailed balance equation between the number densities of triplet and singlet HI, we introduce the rates $\Gamma_{01}^a$ and $\Gamma_{10}^a$ of axion induced conversions $J=0 \rightarrow 1$  and $J=1 \rightarrow 0$ respectively (in the rest of the paper we adopt the units $\bar h= c=k_B=1$),
\be
n_0 \left(C_{01} + P_{01}+ A_{01} N_\gamma + \Gamma_{01}^a\right)
\label{db}
\ee
\[
= n_1 \left(C_{10}+P_{10} + A_{10} (1 +N_\gamma) + \Gamma_{10}^a \right)\,.
 \]
The spin temperature  $T_s$, is defined by the ratio
\be
\frac{n_1}{n_0} = \frac{g_1}{g_0}\, e^{-T_*/T_s}  \simeq 3 \left( 1- \frac{T_*}{T_s} \right)
\ee
where $T_*=E_{21}/k_B=0.068 K$. The rates of collision and spontaneous emissions are related as
\be
C_{01} = \frac{g_1}{g_0} \,C_{10}\, e^{-T_*/T_K} \simeq 3 \, C_{10} \, \left( 1- \frac{T_*}{T_K} \right)
\ee
where $T_K$ is the gas kinetic temperature. The Lyman UV photons induced transitions via the Wouthuysen-Field effect \citep{WF1,WF2}  are determined by the color temperature $T_c$ defined by
\be
P_{01} = \frac{g_1}{g_0} \,P_{10}\, e^{-T_*/T_c} \simeq 3 \, P_{10} \, \left( 1- \frac{T_*}{T_c} \right)
\ee
and $T_c \simeq T_K$. The Einstein spontaneous emission rates $A_{10}$ and $A_{01}$  are related by detailed balance as
\be
A_{01} = \frac{g_1}{g_0} \, A_{10}= 3  \, A_{10}\,.
\ee
The gas temperature $T_K $ and the photon temperature $T_{\gamma}$ are both much larger than $T_*$ . The photon occupation number at 21-cm wavelength is $N_\gamma = (e^{T_*/T_\gamma}-1)^{-1} \simeq T_\gamma/T_*$. Putting these in equation (\ref{db}) we can solve for the spin temperature as a result of collisions with H and CMB photons and with a down conversion by axions with the rate $C_a$,
\be
T_s=  \frac{T_\gamma + x_{col} T_K + x_{\alpha} T_c
+ \frac{T_*}{A_{10} }{\Gamma_{10}^a}
}{1+x_{col} + x_{\alpha}
+\frac{1}{A_{10}}( {\Gamma_{10}^a- \frac{1}{3} \Gamma_{01}^a ) }} \,,
\label{Ts}
\ee
where
\be
x_{col} \equiv \frac{C_{10}}{A_{10}} \frac{T_*}{T_K} \,,\quad x_{\alpha} \equiv \frac{P_{10}}{A_{10}} \frac{T_*}{T_c} \,\,.
\ee
The spontaneous emission rate for photon emission is
\be
A_{10}= \frac{ (2 \pi)^3 \,\alpha \,\nu_{21}^3 }{3 m_e^2}= 2.869 \times 10^{-15} \, s^{-1}=1.88 \times 10^{-30} \,{\rm eV}
\label{A1}
\ee
and that the induced transition rate by CMB interaction is
\be
\gamma=A_{10} N_\gamma \simeq A_{10}\frac{T_\gamma}{T_*} = 1.36 \times 10^{-27}  \left(\frac{z+1}{18} \right) \,{\rm eV}
\label{gamma}
\ee
The brightness contrast between the CMB and the 21-cm absorbing regions at any redshift $z$ is
\be
T_{21}= \frac{T_s-T_\gamma}{1+z} \left(1-e^{-\tau}\right)
\simeq \frac{T_s-T_\gamma}{1+z} \,\tau
\label{t21}
\ee
where the optical depth
\be
\tau= \frac{3}{32 \pi} \frac{T_*}{T_s} \, n_{HI} \lambda_{21}^3\, \frac{A_{10}}{H(z) + (1+z) \partial_r v_r}
\label{tau}
\ee
which depends on the density of HI, the Hubble expansion rate and the velocity gradient of the HI cloud. The brightness contrast temperature is therefore \citep{Madau:1996cs} ,
\be
T_{21}= 26.8\, x_{HI}\, \left(\frac{1+z}{10}  \right)^{1/2} \left( \frac{T_s -T_\gamma}{T_s}\right) \,\,{\rm  mK}
\label{T21}
\ee
where $x_{HI}$ is the neutral hydrogen fraction.
In the standard cosmological models \citep{Barkana:2018lgd} the predicted brightness contrast    $T_{21} \simeq - 200$ mK whereas the observed brightness contrast measured in the EDGES survey \citep{Bowman:2018yin} is $T_{21}= - 500^{+200}_{500}\, $ mK .

The hyperfine splitting of HI arises from the $\vec S_e\cdot \vec S_p$ coupling
 \be
 H_0= E_{10}  \vec S_e\cdot \vec S_p = E_{10} \left( J^2 -S_e^2 -S_p^2 \right) \,, \quad J=0,1\,.
 \ee
 The hyperfine splitting between the $J=0$ and $J=1$ state is \citep{Cohen-Tannoudji:1978},
 \be
 E_{10}= \frac{4}{3} g_p \frac{m_e^2}{m_p} \alpha^4 = 5.9 \mu {\rm eV}.
 \ee
This energy level is ( coincidently )  close to the axion mass which is predicted to arise from the QCD anomaly contribution \citep{diCortona:2015ldu}
\be
m_a= \frac{m_\pi f_\pi }{f_a} \frac{ \sqrt{ m_u m_d}}{m_u+m_d} \simeq 5.7 \mu {\rm eV} \left( \frac{ 10^{12} {\rm \, GeV}}{f_a}\right)
\label{mafa}
\ee
 The QCD axion
\citep{Weinberg:1977ma,Wilczek:1977pj,Kim:1979if,Shifman:1979if,Zhitnitsky:1980tq,Dine:1981rt} has a mass of $m_a= 5.9 \mu eV (10^{12} \text{GeV}/f_a)$ \citep{diCortona:2015ldu} which can serve as dark matter by the coherent oscillation of the axion zero modes \citep{ Preskill:1982cy, Sikivie:2006ni, Bae:2008ue}.
 The dark matter density is related to the axion mass and coupling as
\be
\Omega_a  =0.15 \left( \frac{10^{12} {\rm \, GeV}}{f_a}\right)^{7/6} \theta_1^2
\ee
where $\theta_1$ is the misalignment angle of a axion field in our horizon (if PQ symmetry is broken prior to inflation), which implies that axions with mass $m_a\simeq 5.9 \mu {\rm eV}$ and coupling $f_a\simeq 10^{12} {\rm GeV}$ can account for the dark matter in the universe.

The coupling of axions to fermions at non-relativistic energies is  given by \citep{Sikivie:2014lha}
\be
H_{int}= \frac{g_f}{ f_a} \left( \nabla a(x) \cdot  \vec S + \frac{\partial_t a(x) }{m_f} \,  \vec p_f \cdot \vec S \right)
\label{hint}
\ee
The first term can cause $\Delta L=0$ , $\Delta J=1$ spin-flip transitions and the second term can give rise to $\Delta L=1$ parity changing transitions. The spin flip hyperfine transitions in atoms involve
energy  $\simeq \mu$ eV  while the $\Delta L=1$ transitions of electron orbitals involve energy levels with $eV$ gap and are  therefore not affected by axions bath. We will consider the first term of (\ref{hint})  to study spin flip  para-ortho hydrogen transitions in presence of axions in the $z=15-35$ first-star formation era. Consider the case of axion coupling to the electron, the transition amplitude between  $|J=0, M=0 \rangle$ state and $|J=1, M=0 , \pm 1\rangle$ states is given by
\be
 \frac{g_e}{ f_a} \langle J=1, M=0, \pm 1 | \vec p_a \cdot \vec S_e | J=0, M=0 \rangle\,.
\ee
To compute th the $J=0 \leftrightarrow 1$ transition probability the simplest step is to assume that the electron spin z-axis along the axion momentum $\vec p_a$. We  obtain the following transition amplitudes ,
\bea \label{Sz}
 \langle J=1, M=0 |  \, {S_e}_z \, | J=0, M=0 \rangle= \frac{1}{2}, \\
 \langle J=1, M= \pm 1 |\, {  S_e}_z \, | J=0, M=0 \rangle= 0. \nonumber
\eea
The axion induced spin flip transitions are between the $| J=0,M=0\rangle$ and the $J=1,M=0\rangle$ state of the triplet.
The transition frequency of the $J=0 \leftrightarrow 1$  transitions is therefore
\be
\Omega = \frac{g_e}{2 \, f_a} \, |{\bf p}_a|  \sqrt{n_a(p)}
\label{omega1}
\ee
Now the spin temperature is  defined by the ratio $n_1/n_0 = (g_1/g_0) \exp(-T_s/T_*)$ and $g_1/g_0=3 $ . If there are spin-flip transitions taking place which are dominant then the transition probability for stimulated emission and absorption are equal and the $|J=1, M=0 \rangle$ states are in equilibrium with one $|J=0, M=0 \rangle $ state,  the equilibrium ratio of $J=1$ and $J=0$ states will be $n_1/n_0= 1$. So the prediction for the spin temperature when flip spin processes are dominant is
\be
T_s=  - T_* \, \left(\ln[1/3] \right)^{-1}= 0.062\, {\rm K}.
\ee
Dominant spin flip transitions  will drive the HI spin temperature to $T_s = 0.062\, $K.

In deriving this conclusion we used the fact that only one state from the three degenerate states of $J=1$ mix with the $J=0$ state due to the spin-flip process. This would remain true even if we take the spin z-axis of the electron in any general direction w.r.t the  axion momentum. In that case the transition is between the $|J=0,M=0\rangle$ state and   one linear combination of the
three $|J=1, M=0,\pm1\rangle$ states. We can see that by calculating  the state into which the singlet state is transformed into by the axion operator,
\bea
\vec S_e \cdot \hat p_a \, | 0,0 \rangle &=&   \frac{ {\hat p}_z }{2} \,|1,0 \rangle
 + \frac{- {\hat p}_x+i {\hat p}_y}{2 \sqrt{2}} \,|1,1 \rangle + \frac{ {\hat p}_x+i {\hat p}_y}{2 \sqrt{2}} \,|1,-1 \rangle  \nonumber\\
&\equiv&  \frac{1}{2 } \left( \alpha_1 |1,0 \rangle + \alpha_2 |1,1\rangle +\alpha_3 |1, -1\rangle\, \right)
\label{alpha}
\eea
with $\alpha_1^2+\alpha_2^2+\alpha_3^2=1$. The transition will take place between the singlet state and the linear combination of the triplet states given in the r.h.s of (\ref{alpha}) with
 the oscillation frequency given by (\ref{omega1}). In the rest of the paper we will take the spin z-axis of the electron along the axion momentum with no loss of generality.

The coherent oscillation mechanism will be  effective as long as the time scale of all other processes (like photon induced transitions between the levels) are slower than the axion induced transition frequency $\Omega \Gamma_\gamma < 1$.


We see that only one linear combination of $J=1$ states mixes with the $J=0$ state due to the axion process.

This is different compared to the photon absorption/emission processes which are incoherent and connect all the three states of $J=1$ to the $J=0$ state. The foregoing analysis would work identically if the axion couples to nucleons and the HI $J=1 \leftrightarrow J=0$ transitions take place by spin flip of the proton by axion emission/absorption since unlike the photon dipole moment the axion coupling is not suppressed by fermion mass \citep{Auriol:2018ovo}. In KSVZ models \citep{Kim:1979if, Shifman:1979if}  the axion couples to quarks but not leptons and $g_f^2= g_p^2 \sim 0.22$ while in the DFSZ \citep{Dine:1981rt, Zhitnitsky:1980tq} models the axion couples to both quarks and leptons and $g_f^2= (g_p-g_e)^2 \sim 0.3$.

Taking into account all the interactions which determine the spin temperature (\ref{Ts}),  we set the axion-induced rates $\Gamma^a_{01}=\Gamma^a_{10}$, so the
the spin temperature reduces to
\be
T_s = \frac{ A_{10} T_\gamma+  C_{10}  T_* + P_{10}T_* + \Gamma^a_{10}T_*} {A_{10}+ C_{10} \frac{T_*}{T_k} + P_{10} \frac{T_*}{T_c} + \frac{2}{3} \Gamma^a_{10}}
\label{Ts1}
\ee
If the axion induced transitions $\Gamma_{10}^a$ were to be the dominant $1\leftrightarrow 0$ process then  the spin temperature (\ref{Ts1}) $T_{s} \rightarrow ( 3/2) T_*=0.1 {\rm K}$.
This goes in the right direction of the EDGES observation of $T_{21}= -500^{+200}_{-500} $ mK
which from (\ref{T21}) implies  a spin temperature $T_s= 3.26^{1.94}_{-1.58} $K \citep{Barkana:2018qrx} whereas the prediction from standard astrophysics is $T_a \geq T_k \simeq 6.8$K. Therefore an axion induced spin-flip mechanism can lower the spin temperature and can be a viable explanation of the EDGES observation (we disagree with the conclusion of \citep{Auriol:2018ovo} that spin flip by axion absorption/emission if dominant would take the spin temperature to infinity).

\section{Spin flip by incoherent emission and absorption in the axion bath }


The calculation of the spin flip by axion emission or absorption in the incoherent process is done in several ways in  \citep{Auriol:2018ovo}. We summarize the main steps to highlight the difference between the incoherent and the coherent process we will discuss in the next section.

The matrix element for the atomic transitions is
\be
\langle J=1, M=0| H_{int} |J=0,M=0 \rangle= \frac{g_f}{2 f_a} \nabla a(x)
\ee
The transition amplitude for the $1\rightarrow 0$ process by stimulated emission is
\bea
{\cal M}_{10}
= \frac{g_f}{2 f_a} |p_a| \frac{1}{\sqrt{2 E_a V}} \sqrt{N_p+1}
\eea
The decay width of the atom is therefore
\bea
\Gamma_{10}^a &=& \left(\frac{g_f}{2 f_a} \right)^2 V \int \frac{d^3 p }{(2 \pi) } \delta (E_a -E_{10}) p_a^2 \frac{1}{2 E_a V} (N_p+1)\nonumber\\
&=& \frac{1}{16 \pi^2} \left(\frac{g_f}{ f_a} \right)^2  p_{10}^3 (N_p+1)
\eea
where $p_{10}= \sqrt{E_{10}^2-m_a^2}$.

Similarly the transition amplitude for stimulated absorption is
\bea
{\cal M}_{01}
= \frac{g_f}{2 f_a} |p_a| \frac{1}{\sqrt{2 E_a V}} \sqrt{N_p}
\eea
and the rate for stimulated absorption is
\be
\Gamma_{10}^a = \frac{1}{16 \pi^2} \left(\frac{g_f}{ f_a} \right)^2  p_{10}^3  N_p.
\label{Gamma10}
\ee
Here $N_p$ is the occupation number  (not number density)  of axions with momenta $p_{10}$.
In case of axions $N_p \gg 1$ and the rates for stimulated emission and absorption are same, $\Gamma^a_{10}= \Gamma^a_{01}$.
The rate for spontaneous axion emission is
\be
A_{10}^a=\frac{1}{16 \pi^2} \left(\frac{g_f}{ f_a} \right)^2  p_{10}^3
\ee

The numerical value of $\Gamma^{a}_{10}$ (\ref{Gamma10})  is
\be
\Gamma^a_{10}= 6.2  \times 10^{-27} g_f^2\, p_{10}^3 \,N_p \quad {\rm GeV^{-2}}
\label{Gamma-1}
\ee

Axion dark matter are a Bose-Einstein condensate with de-Broglie wavelength $\lambda_a= 2\pi/(m_a v_a)$. The occupation number of the  axion Bose condensate is related to their number density as
\be
N_0= n_a \lambda_a^3 = \left(\frac{\rho_{DM}(z)}{m_a}\right) \left( \frac{2 \pi}{m_a v_a} \right)^3 =
\left(\frac{\rho_{DM}}{m_a}\right) \left( 2 \pi t(z) \right)^3\,
\label{N0}
\ee
where the axion rms velocity $v=(m_a t )^{-1}$ \citep{Sikivie:2009qn}. At $z=18$ is the zero mode occupation number of axions is $N_0(z \simeq17)= 10^{63}$. The combination $p^3 N_p= (2 \pi)^3 n_p= (2\pi)^3 (\rho_a/m_a) = 4.5 \times 10^{-31} \,{\rm Gev^3}$. Therefore the axion induced transition rate $\Gamma^a_{10} \simeq 10^{-46} \,{\rm eV}$.
This is true even if the mass is exactly equal to $E_{21}$  (this can happen if the axion mass varies with time  during this epoch as we shall discuss in the next section) and $N_p=N_0$. So in conclusion the incoherent axion absorption/emission always is much lower by many orders of magnitude compared to the photon process.

 In the next section we will examine a coherent axion emission/absorption process which leads to spin oscillation in HI and the is proportional to $1/f_a$.




\section{ Spin oscillations by coherent absorption/emission of axions - the Jaynes-Cummings model }

The axion operators $a$ and $a^\dagger$ will have non-zero matrix elements between $|N_p \rangle$ and $|N_p+1\rangle$ occupation number states. The Hamiltonian of the two-level atom coupled to an axion background can be thus written in the $ |\Psi_0 \rangle = |1,0\rangle \otimes |N_p+1 \rangle$ and $ |\Psi_1 \rangle= |1,1\rangle \otimes |N_p\rangle$ basis can be expressed as
\be
H= E_{10} \,\frac{1}{2} \sigma_z + E_a a^\dagger a + \lambda \left( \sigma^+\, a + \sigma^- a^\dagger \right) \label{JCa2}
\ee
\[
\quad=\left(
\begin{array}{ccc}
 \frac{1}{2} E_{10}  + E_a\, N_p &\lambda (N_p+1)^{1/2}     \\
 \lambda (N_p+1)^{1/2}  & -\frac{1}{2} E_{10} + E_a\, (N_p+1)   &
\end{array}
\right)
\]
where  the off-diagonal terms in the Hamiltonian are
\be
\langle \Psi_1 |H_{int}| \Psi_0 \rangle = \lambda (N_p+1)^{1/2}
\ee
and $\lambda= \frac{g_f p_a}{2 f_a \sqrt{2 E_a V}}$. This Hamiltonian is identical to that of the Jaynes-Cummings system of a two level atom in a cavity ( Appendix \ref{JCsec}).
We can  simplify (\ref{JCa2}) by subtracting a term proportional to the identity matrix $(N_p  -1 /2)E_a {\mathcal I}$ and write the Hamiltonian responsible for the transitions between the two levels as
 \[
  H=\left(
\begin{array}{ccc}
 \frac{1}{2}( E_{10}  - E_a)  &\Omega     \\
 \Omega & -\frac{1}{2} (E_{10} - E_a)   &
\end{array}
\right)
\]
\label{JCa3}
where the frequency of transitions between the two states
\bea
\Omega&=&\lambda (N_p+1)^{1/2} = \frac{g_f p_a}{2 f_a \sqrt{2 E_a }} \left( \frac{N_p+1}{V} \right)^{1/2} \nonumber\\
&=& \frac{g_f }{2  f_a \sqrt{2 E_{10}}   }  \left(E^2_{10}-m^2_a \right)^{1/2} \sqrt{ n_p}
\label{Omega}
\eea
where we have written $(N_p +1)/V \simeq  N_p/V =n_p $ which is  the  number density of axions with momentum $p=p_{10}= \sqrt{E^2_{10}- m^2_a}$.

There is a mixing of the two levels with the mixing angle given by
\be
\theta= \frac{1}{2}\tan^{-1} \left(\frac{2 \Omega }{E_{10} -E_a}\right)
\ee
and the probability of $1\leftrightarrow 0$ transitions is then
\be
P_{10}^a(t) = \frac{\Omega^2}{(E_{10} -E_a)^2 + \Omega^2} \, \sin \left[ \sqrt{(E_{10} -E_a)^2 + \Omega^2} \,\,\, t \right] \,.
\ee
The coherent axion emission/absorption process will proceed for a time limited by the lifetime of the levels due to photon emission-absorption and collisions (which are incoherent processes). If the line width of these other processes is $\gamma $  (\ref{gamma})  then time averaged oscillation probability between the states is \citep{Cohen-Tannoudji:1978}
\be
\bar P_{10}^a= \int_0^{\infty} \,\, dt e^{-\gamma t } \,P_{10}^a(t)
\ee
which gives us the transition probability between the $1\leftrightarrow 0$ levels by the axion spin flip mechanism as
\be
\bar P_{10}^a= \frac{1}{2} \frac{\Omega^2}{\gamma^2 + \Omega^2 + (E_{10} -E_a)^2 }
\label{barp1}
\ee
since $E_a=\sqrt{p_a^2 +m_a^2}= E_{21}$ by emission/absorption an on-shell axion  of a fixed momentum $p_a=p_{10}$. We have the transition probability (\ref{barp1}) reduces to the resonant  case
\be
\bar P_{10}^a= \frac{1}{2} \frac{\Omega^2}{\gamma^2 + \Omega^2 }
\label{barp2}
\ee
The up and down transitions have the same probability $\Gamma^a_{01}/\Gamma^a_{10}=1$ as assumed in equation (\ref{Ts1}).
If the axion induced transition is dominant over
 to the photon processes i.e $\Omega \geq \gamma$  we will have $\bar P_{10}^a= \frac{1}{2}$.  The transition rate for the up or down conversions in equation (\ref{Ts1}) will be
 $\Gamma^a_{10}= \Omega \bar P_{10}^a$.

We estimate the value of the axion induced transition frequency as follows, from  (\ref{Omega}) we get
\be
\Omega = 0.77 \times 10^{-24}\frac{ \sqrt{n_p}}{\rm eV^{3/2}}\,\, {\rm eV}\, \left(\frac{\sqrt{E_{10}^2-m_a^2}}{0.9 E_{10}}\right)
\left(\frac{g_f \, 10^{12} {\rm GeV}}{f_a} \right).
\label{Omeganp}
\ee
For this calculation we have taken $E_{10} $ and $m_a$ to differ such that $p_a=0.9 E_{10}$ and the axions absorbed/emitted are relativistic. In order that the axion induced transition to occur at the same rate as the photon induced transitions $\gamma =1.36 \times 10^{-27} {\rm eV}$ (\ref{gamma}) we need axion number density $n_p \simeq 10^{-6} {\rm eV}^3$ in a coherent state with momenta $\langle p_a \rangle \sim  6 \times 10^{-6}  {\rm eV}$. For comparison we note that   the number density of thermal axions is $n_a(T=16 K)= 0.45 \times 10^{-3} {\rm eV^3}$ \citep{Masso:2002np}.

\section{Coherent axion production from cosmic string decay}

In the this section we discuss a mechanism whereby a coherent background of relativistic axions peaked at a particle frequency  can be obtained by the decay of cosmic strings following references \citep{ Yamaguchi:1998gx, Hiramatsu:2010yu, Gorghetto:2018myk}. The
peak energy of the axions will depend upon the Horizon size at the time of decay. We will find the time of decay of the cosmic string which would  produce axions peaked at the frequency $6 \times 10^{-6} {\rm eV}$ which we require for the spin-flip HI transitions discussed in the previous section.

Consider a complex scalar field with a global $U(1)$ symmetry. Its potential  will be of the general form
\be
V(\phi) = \frac{\lambda}{2} \left( (\phi^\dagger \phi)^2 - \frac{f_a}{2} \right) + \frac{\lambda}{3} T^2  \phi^\dagger \phi
\ee
where the second term is the finite temperature correction. At high temperatures ( $T > T_c= \sqrt{3/2} f_a$ ) minima of the potential is at $|\phi| =0$ and the $U(1)$ symmetry is unbroken. A the temperature goes below the critical temperature $T<T_c$ , the minimum of the potential shifts to $|\phi_{min}| = (f_a/\sqrt{2}) (1-T/T_c)^{1/2}$. During the first order phase transition a one dimensional topological defect is produced. In the broken symmetric phase we can express the complex scalar as
\be
\phi(x) = \frac{1}{\sqrt{2}} \left(f_a + \rho(x) \right) e^{i \frac{a(x)}{f_a}}
\ee

The mass of the radial field  $\rho(x)$ is  $m_\rho  =\sqrt{ \lambda} f_a$  while the axion field $a(x)$ is massless and squires a mass after the QCD phase transition. The cosmic string is a cylindrical configuration of the false vacuum $|\phi|=0$ and surrounded by the true vacuum $|\phi_{min}| =f_a/ \sqrt{2}$. The cosmic string network forms a stable configuration and the strings decay over cosmological timescales into axions. The peak axion spectrum has a sharp peak in momentum which which is of size of Horizon at the time of decay. In the radiation era $t=1/(2H)$ and the mean value of the frequency as a function of time of decay obtained from lattice simulations  \citep{ Yamaguchi:1998gx,Hiramatsu:2010yu} is
\be
\langle k^{-1} \rangle = \epsilon^{-1} \frac{t}{2 \pi}\,. \quad \epsilon^{-1}= 0.25 \pm 0.18\,.
\ee
If the peak frequency is to be $k=6 \times 10^{-6} {\rm eV}$ the temperature is $100 \,{\rm GeV}$ . The axion number is ( dropping log corrections \citep{Yamaguchi:1998gx,Hiramatsu:2010yu} ),
\bea
n_a(k) &\simeq& \frac{f_a^2 }{\epsilon t} = \frac{f_a^2} {2 \pi} \frac{1}{ \epsilon^2 \, \langle k^{-1} \rangle}\nonumber\\
&\sim & 10^{36}\, {\rm eV^3}\, \left( \frac{f_a}{10^{12} {\rm GeV}}\right)^2  \left(\frac{k}{ 6 \times 10^{-6} \,{\rm eV}} \right)\,.
\eea
The number density of the radiated axions dilutes with the scale factor  as $ n_a(T) = R(T)^{-3}$.
The axion number at the cosmic dawn epoch $T=17^{\circ} K$ is
\be
n_a = 10^{-5} \, {\rm eV^3}\,.
\ee
This is larger than the number density $10^{-6} \, {\rm eV^3}$ required for the axion induced spin flip rate (\ref{Omeganp})  to dominate over the photon induced spin flip rate (\ref{gamma}). The contribution of this population of axions in the present universe energy density is $\rho_a= m_a n_a = (7 \times 10^{-4} {\rm eV})^4  $ and it makes a subdominant contribution to the critical density in the present universe $\rho_{c}= (2.24 \times 10^{-3} {\rm eV})^4$. If one integrates over the total axion emission over all $t$ then the axions missed by cosmic strings have the potential to make a substantial contribution to dark matter in the present universe \citep{ Yamaguchi:1998gx, Hiramatsu:2010yu, Gorghetto:2018myk} .

\section{Conclusions}

Motivated by the fact that the axion mass is close to the energy  of the HI singlet-triplet 21-cm transition, we show that axion emission or absorption induced by the background dark matter axions can lower the spin temperature of HI sufficiently to remove the discrepancy between the prediction of the brightness contrast  temperature in  standard cosmological models  and the EDGES observation \citep{Bowman:2018yin}. The resonant emission condition requires that the energy of the emitted axion  should be equal to $ E_{10}$ to within the line width $\gamma$. This causes an  induced emission/absorption of on shell axions of momentum $p_a=(E_{10}^2-m_a^2)^{1/2}$. Noting the  the equivalence  with the  Jaynes-Cummings model we develop  the equations for spin-flip of HI as a Rabi-like  oscillation in the axion background. The important difference with  Rabi oscillations in a classical field background is that there for resonance oscillations the classical background frequency must match the energy difference. Here in  the Jaynes-Cummings treatment where the background field is also quantized  the  momentum of the emitted/absorbed  on-shell axion adjusts to ensure that the resonance condition $E_{10}=E_a$ always met. In the coherent transition process the transition frequency $\Omega \propto 1/f_a$ as opposed to the incoherent emission/absorption process where the decay with is width $\Gamma \propto 1/f_a^2$. The hydrogen has a velocity dispersion so one must be sure that the relation $E_{10}- E_a < \Omega$ is maintained despite the motion of the hydrogen. This relation is valid as the axion energy emitted from cosmic strings  is not one sharp line but a broad-band spectrum. So despite the fact that $E_a$ will have a small spread due to the non-relativistic motion of hydrogen,  if the energy width of the emitted axions is larger than the energy smearing due to hydrogen motion, the atom will absorb and emit energy equal to $E_{10}$ and the resonance condition  $E_{10}- E_a < \Omega$  will be satisfied.

For the specific application to the case of the 21-cm brightness temperature we show that the axionic spin flip processes when dominant can take the spin temperature to $T_s \rightarrow 0.1 $K compared to the standard astrophysics where the spin temperature $T_s  \geq T_k \simeq 6.8 $K.  The coherent background of relativistic axions emitted by the decay of cosmic strings which have survived to the electroweak era can explain the lower than observed HI spin temperature. The implication of  this mechanism is that  the spin temperature of HI is $T_{s}= (5.2-1.68)$K which can be tested in future 21-cm expeiments.



\section*{Acknowledgements}

G.L. and S.M. thank INFN for support.




\bibliographystyle{mnras}

\begin{thebibliography}{99}


\bibitem [\protect\citeauthoryear{Auriol}{2018}]{Auriol:2018ovo}
  Auriol, A., Davidson, S., Raffelt, G., 2019, Phys. Rev. D, 99, 023013


%

\bibitem [\protect\citeauthoryear{Bae}{2008}]{Bae:2008ue}
  Bae, K. J., Huh, J. H., Kim, J. E., 2008, JCAP, 0809, 005

\bibitem[\protect\citeauthoryear{Barkana}{2018}]{Barkana:2018lgd}
  Barkana, R., 2018, Nature, 555, 71

\bibitem[\protect\citeauthoryear{Barkana et al.}{2018}]{Barkana:2018qrx}
  Barkana, R., Outmezguine, N. J., Redigolo,D., Volansky, T., 2018, Phys. Rev. D, 98, 103005

\bibitem[\protect\citeauthoryear{Berlin et al.}{2018}]{Berlin:2018sjs}
  Berlin, A., Hooper,D., Krnjaic, G., McDermott,S.~D., 2018, Phys. Rev. Lett. 121, 011102






\bibitem[\protect\citeauthoryear{Bowman et al.}{2018}]{Bowman:2018yin}
  Bowman, J.~D., Rogers, A.~E.~E., Monsalve,R.~A., Mozdzen,T.~J., Mahesh, N., 2018, Nature, 555, 67

\bibitem[\protect\citeauthoryear{Cohen-Tannoudji}{1978}]{Cohen-Tannoudji:1978}
Cohen-Tannoudji,C., Diu,  B., Laloe, F.,  1978, Quantum Mechanics, Vol. 2,  Publisher: Wiley

\bibitem[\protect\citeauthoryear{D'Amico et al.}{2018}]{DAmico:2018sxd}
  D'Amico,G., Panci,P., Strumia, A., 2018, Phys. Rev. Lett., 121, 011103

\bibitem[\protect\citeauthoryear{DEramo}{2018}]{DEramo:2018vss}
  D'Eramo, F., Ferreira, R. Z., Notari, A., Bernal, J. L., 2018, JCAP, 1811, 014

\bibitem[\protect\citeauthoryear{Dine}{1981}]{Dine:1981rt}
  Dine, M., Fischler,W., Srednicki, M., 1981, Phys. Lett., 104B, 199

\bibitem[\protect\citeauthoryear{Du et al.}{2018}]{Du:2018uak}
  Du,N., {\it et al.} [ADMX Collaboration], 2018, Phys. Rev. Lett., 120, 151301

\bibitem[\protect\citeauthoryear{Dvorkin et al}{2014}]{Dvorkin:2013cea}
  Dvorkin, C., Blum, K., Kamionkowski, M., 2014, Phys. Rev. D, 89, 023519


\bibitem[\protect\citeauthoryear{Erken et al.}{2012}]{Erken:2011dz}
  Erken, O., Sikivie, P., Tam,H., Yang, Q., 2012, Phys. Rev. D, 85, 063520

\bibitem[\protect\citeauthoryear{Fialkov et al.}{2018}]{Fialkov:2018xre}
  Fialkov, A., Barkana,R., Cohen, A., 2018, Phys. Rev. Lett., 121, 011101

\bibitem[\protect\citeauthoryear{Field}{1958}]{WF2}  G. B. Field, 1958, Proc. I.R.E. 46 240

\bibitem[\protect\citeauthoryear{Fraser et al.}{2018}]{Fraser:2018acy}
  Fraser, S., Hektor, A., Hutsi, G., Kannike, K., Marzo, C., Marzola, L.,  Spethmann, C., Racioppi, A., Raidal, M., Vaskonen, V.,  Veermae, H.,
  2018, Phys. Lett. B, 785, 159


\bibitem[\protect\citeauthoryear{Furlanett et al.}{2006}]{Furlanetto:2006jb}
  Furlanetto, S., Oh, S.P., Briggs, F., 2006, Phys. Rept., 433, 181

\bibitem[\protect\citeauthoryear{Gerry}{2005}]{Gerry:2005}
 Gerry, C., Knight, P., L., 2005, Introductory Quantum Optics, Cambridge University Press

\bibitem[\protect\citeauthoryear{Gorghetto}{2018}]{Gorghetto:2018myk}
  Gorghetto, M., Hardy,E., Villadoro, G., 2018, JHEP, 1807, 151

\bibitem[\protect\citeauthoryear{Grilli di Cortona et al.}{2016}]{diCortona:2015ldu}
  Grilli di Cortona, G., Hardy, E., Pardo Vega, J., Villadoro, G., 2016, JHEP, 1601, 034

\bibitem[\protect\citeauthoryear{Hiramatsu}{2011}]{Hiramatsu:2010yu}
  Hiramatsu, T., Kawasaki, M., Sekiguchi, T., Yamaguchi, M., Yokoyama, J., 2011, Phys. Rev. D, 83, 123531


\bibitem[\protect\citeauthoryear{Houston et al.}{2018a}]{Houston:2018vrf}
  Houston, N., Li, C., Li, Yang,T. Q., Zhang, X., 2018, Phys. Rev. Lett., 121, 111301

\bibitem [\protect\citeauthoryear{Houston et al.}{2018b}] {Houston:2018vbk}
  Houston, N., Li, C., Li,  T., Yang, Q., Zhang, X., arXiv:1812.03931 [hep-ph]

\bibitem[\protect\citeauthoryear{Jaynes}{1963}]{Jaynes:1963}
    Jaynes,E., Cummings, F. W., 1963. Proceedings of the IEEE, 51, 89

\bibitem[\protect\citeauthoryear{Kim}{1979}]{Kim:1979if}
  Kim, J. E., 1979, Phys. Rev. Lett., 43, 103


\bibitem [\protect\citeauthoryear{Masso}{2002}]{Masso:2002np}
  Masso, E., Rota, F., Zsembinszki, G., 2002, Phys. Rev. D, 66, 023004

\bibitem[\protect\citeauthoryear{Madau et al.}{1997}]{Madau:1996cs}
  Madau, P., Meiksin, A., Rees, M.J., 1997, Astrophys. J., 475, 429

\bibitem[\protect\citeauthoryear{Moroi et al.}{2018}]{Moroi:2018vci}
  Moroi, T., Nakayama, K., Tang, Y., 2018, Phys. Lett B, 783, 301

\bibitem[\protect\citeauthoryear{Munoz and Loeb}{2018}]{Munoz:2018pzp}
  Munoz, J. B., Loeb, A., 2018, Nature, 557, 684

\bibitem[\protect\citeauthoryear{Munoz et al.}{2018}]{Munoz:2018jwq}
  Munoz, J. B., Dvorkin, C., Loeb, A., 2018, Phys. Rev. Lett., 121, 121301

\bibitem[\protect\citeauthoryear{Preskill}{1983}]{Preskill:1982cy}
  Preskill, J., Wise, M. B., Wilczek, F., 1983, Phys. Lett. B,120, 127

\bibitem[\protect\citeauthoryear{Pospelov et al.}{2018}]{Pospelov:2018kdh}
  Pospelov, M., Pradler, J., Ruderman, J. T., Urbano, A., 2018, Phys. Rev. Lett., 121, 031103


\bibitem[\protect\citeauthoryear{Shifman}{1980}]{Shifman:1979if}
  Shifman, M. A., Vainshtein, A. I., Zakharov, V. I., 1980,  Nucl. Phys. B, 166, 493

\bibitem[\protect\citeauthoryear{Shore}{1993}]{Shore:1993}
  Shore, B. W., Knight, P. L., 1993, Journal of Modern optics, 40, 1195


\bibitem [\protect\citeauthoryear{Sikivie}{2008}] {Sikivie:2006ni}
  Sikivie, P., 2008, Lect. Notes Phys. 741, 19

\bibitem[\protect\citeauthoryear{Sikivie and Yang}{2009}]{Sikivie:2009qn}
  Sikivie, P., Yang, Q., 2009, Phys. Rev. Lett.,  103, 111301

\bibitem[\protect\citeauthoryear{Sikivie}{2014}]{Sikivie:2014lha}
  Sikivie, P., 2014, Phys. Rev. Lett., 113, 201301



\bibitem[\protect\citeauthoryear{Weinberg}{1977}]{Weinberg:1977ma}
  Weinberg, S., 1978, Phys. Rev. Lett., 40, 223

\bibitem[\protect\citeauthoryear{Wilczek}{1977}]{Wilczek:1977pj}
  Wilczek, F., 1978, Phys. Rev. Lett., 40, 279

\bibitem[\protect\citeauthoryear{Wouthuysen}{1952}]{WF1}  Wouthuysen, S. A., 1952, Astron. J.,  57, 31


\bibitem[\protect\citeauthoryear{Yamaguchi}{1999}]{Yamaguchi:1998gx}
  Yamaguchi, M., Kawasaki, M., Yokoyama, J., 1999, Phys. Rev. Lett., 82, 4578



\bibitem[\protect\citeauthoryear{Zhitnitsky}{1980}]{Zhitnitsky:1980tq}
  Zhitnitsky, A. R., 1980,  Sov. J. Nucl. Phys., 31, 260

%



  \end{thebibliography}




\appendix
\section{Jaynes-Cummings model }  \label{JCsec}
The Jaynes-Cummings model represents the interaction of a two level atom -states $|g\rangle$ and $|e\rangle$ which can absorb or emit bosons from a large particle number bosonic state. The Hamiltonian in the  \citep{ Jaynes:1963, Shore:1993, Gerry:2005},
\be
H= \frac{\omega_0}{2} \sigma_z + \omega a^\dagger a + \lambda \left( \sigma^+ a + \sigma^- a^\dagger \right)
\label{JC}
\ee
the Pauli matrices act on the two level atomic states and the creation (annihilation) operators $a (a^\dagger)$ operate on the n-particle bosonic states $|n\rangle$. The energy level of the atoms are differ by $\omega_0$, the energy of the bosons is $\omega$ and $\lambda$ represents the coupling constant for the transitions between  the two atomic $|g\rangle=(0,1)^T$ and $|e\rangle=(1,0)^T$ levels by stimulated absorption  or emission of a boson into the bath. The Hamiltonian connects the two sets of states $|\psi^+\rangle=|e, n\rangle$ and $|\psi^- \rangle = |g, n+1\rangle$  which in two component notation can be written as
\[
|\psi^+\rangle = \left(\begin{array}{ccc}
|n\rangle\\
0
\end{array}
\right)
\,, \quad |\psi^-\rangle = \left(\begin{array}{ccc}
0\\
|n+1\rangle
\end{array}
\right)
\]
and in this basis the Hamiltonian (\ref{JC}) can be written as
 \[
H=\left(
\begin{array}{ccc}
 \frac{1}{2}\omega_0 + \omega  &\lambda (n+1)^{1/2}     \\
 \lambda (n+1)^{1/2}  & -\frac{1}{2} \omega_0 + \omega (n+1)   &
\end{array}
\right)
\]
The eigenvalues of this Hamiltonian are
\be
E_\pm= \left(n+\frac{1}{2} \right) \omega \pm \Omega,
\ee
where
\be
\Omega\equiv \left( (\omega_0 -\omega)^2 + 4 \lambda^2 (n+1)\right)^{1/2}
\ee
is the Rabi frequency of oscillations between the two states.
The eigenstates of the Hamiltonian are
\bea
|+ \rangle&=& \cos \theta  |\psi^+ \rangle + \sin \theta  |\psi^- \rangle\nonumber\\
|- \rangle&=&\!\! \!\!-\sin \theta  |\psi^+ \rangle + \cos \theta |\psi^- \rangle
\eea
where the mixing between the two states is given by
\be
2\theta=\tan^{-1} \left(\frac{2 \lambda \sqrt{n+1}}{\omega_0 -\omega}\right)
\ee
The mixing is maximal $\theta=\pi/4$ when $\omega_0 -\omega \ll 2 \lambda \sqrt{n+1}$ and the oscillation between the two levels
takes place with the frequency $\Omega= 2 \lambda \sqrt{n+1}$.

The Jaynes-Cummings model of the two level system has also been studied assuming the photons in the cavity to be in a coherent state $\alpha\rangle$ or in a thermal distribution \citep{Gerry:2005}. The oscillation probabilities then have to averaged over the initial and final states which are line combinations of $|n\rangle$ for the case of $\alpha\rangle$ state photons and treated in the density matrix formulation for the case of thermal photons.




\bsp	
\label{lastpage}
\end{document}